\documentclass[preprint,double-spaced,showpacs,amsmath,amssymb]{revtex4}
\usepackage{graphicx}% Include figure files
\usepackage{dcolumn}% Align table columns on decimal point

\begin{document}

\newcommand{\beq}{\begin{equation}}
\newcommand{\eeq}{\end{equation}}
\newcommand{\bea}{\begin{eqnarray}}
\newcommand{\eea}{\end{eqnarray}}
\newcommand{\eps}{\varepsilon}
\newcommand{\Fs}{\mbox{\scriptsize F}}
\newcommand{\pr}{_{\perp}}
\newcommand{\bk}{{\bf k}}
\newcommand{\bs}{{\bf s}}
\newcommand{\bp}{{\bf p}}
\newcommand{\br}{{\bf r}}
\newcommand{\bR}{{\bf R}}
\newcommand{\lsim}{\stackrel{\scriptstyle <}{\phantom{}_{\sim}}}
\newcommand{\gsim}{\stackrel{\scriptstyle >}{\phantom{}_{\sim}}}
\newcommand{\Vef}{V_{\mbox{\scriptsize eff}}}

\newcommand{\VF}{{\mathcal V}^{\mbox{\scriptsize F}}_{\mbox{\scriptsize eff}}}

\title{ Semi-microscopic model of pairing in nuclei}

\author{ S.\,S. Pankratov}
\email{pankratov@mbslab.kiae.ru} \affiliation{Kurchatov Institute,
123182, Moscow, Russia} \affiliation{Moscow Institute of Physics and
Technology, 123098 Moscow, Russia.}

\author{ M. Baldo}
\email{baldo@ct.infn.it} \affiliation{INFN, Sezione di Catania, 64
Via S.-Sofia, I-95125 Catania, Italy}

\author{ U. Lombardo}
\email{lombardo@lns.infn.it} \affiliation{INFN-LNS and University
of Catania, 44 Via S.-Sofia, I-95125 Catania, Italy}

\author{ E.\,E. Saperstein}
\email{saper@mbslab.kiae.ru}
\affiliation{Kurchatov Institute, 123182, Moscow, Russia}

\author{ M.\,V. Zverev}
\email{zverev@mbslab.kiae.ru} \affiliation{Kurchatov Institute,
123182, Moscow, Russia} \affiliation{Moscow Institute of Physics and
Technology, 123098 Moscow, Russia.}

\date{\today}

\begin{abstract}
A semi-microscopic model for nucleon pairing in nuclei is presented starting from the {\it ab intio} BCS gap
equation with Argonne v$_{18}$ force and the self-consistent Energy Density Functional Method basis characterized
with the bare nucleon mass. The BCS theory is formulated in terms of  the model space $S_0$ with the effective
pairing interaction calculated from the first principles in the subsidiary space $S'$. This effective interaction
is supplemented with a small phenomenological addendum containing one phenomenological parameter  universal for
all medium and heavy atomic nuclei. We consider the latter as a phenomenological way to take into account both the
many-body corrections to the BCS theory and the effective mass effects. For protons, the Coulomb interaction is
introduced directly. Calculations made for several isotopic and isotonic chains of semi-magic nuclei confirm the
validity of the model. The role of the self-consistent basis is stressed.

\end{abstract}


\maketitle

\section{Introduction}

   In the last few years, some progress has been made in the microscopic
theory of  pairing in nuclei by the Milan group
\cite{milan1,milan2,milan3} and by Duguet et al. \cite{Dug1,Dug2}.
In the first paper of the Milan series, the Bardeen-Cooper-Shrieffer
 (BCS) gap equation for neutrons with the
Argonne v$_{14}$ potential was solved for the nucleus $^{120}$Sn which, being located in the middle of the tin
isotopic chain, is a traditional benchmark for the nuclear pairing problem. The Saxon-Woods Shell-Model basis with
the bare neutron mass $m^*=m$ was used, and the discretization method in a spherical box was applied to take into
account the continuum states restricted by the limiting energy $E_{\rm max}=600$ MeV. Rather optimistic results
were obtained for the gap value, $\Delta_{\rm BCS}=2.2$ MeV. Although it is bigger of the experimental one,
$\Delta_{\rm exp}\simeq 1.3$ MeV, the difference is not so dramatic and left the hope to achieve a good agreement
by developing corrections to the scheme. In Refs. \cite{milan2,milan3} the basis was enlarged, i.e. $E_{\rm
max}=800$ MeV, and, what is more important, the effective mass $m^*\neq m$ was introduced into the gap equation.
The new basis was calculated within the Skyrme--Hartree--Fock (SHF) method with the Sly4 force \cite{SLy4}, that
makes the effective mass $m^*(r)$ coordinate dependent, remarkably different from the bare one $m$. E.g., in
nuclear matter the Sly4 effective mass is equal to $m^*=0.7 m$. As it is well known, in the so called weak
coupling limit of the BCS theory, the gap is exponentially dependent,i.e. $\Delta \propto \exp(1/g)$, on the
inverse dimensionless pairing strength $g= m^*\Vef k_{\rm F}/\pi^2$, where $\Vef$ is the effective pairing
interaction. Therefore, a strong suppression of the gap takes place in the case of $m^*< m$. The value of
$\Delta_{\rm BCS}=0.7$ MeV was obtained in Ref. \cite{milan2} and $\Delta_{\rm BCS}=1.04$ MeV, in Ref.
\cite{milan3}. In both cases, the too small value of the gap was explained by invoking various many-body
corrections to the BCS approximation. The main correction is due to the exchange of low-lying surface vibrations
(``phonons''), contributing to the gap around 0.7 MeV \cite{milan2}, so that the sum  turns out to be $\Delta=1.4$
MeV very close to the experimental value. In \cite{milan3} the contribution of the induced interaction caused by
exchange of the high-lying in-volume excitations was added either, the sum again is equal to $\Delta\simeq 1.4$
MeV. Thus, the calculations of Refs. \cite{milan2,milan3} showed that the effects of $m^*\neq m$ and of many-body
corrections to the BCS theory are necessary  to explain the difference of ($\Delta_{\rm BCS}-\Delta_{\rm exp}$).
In addition, they are of different sign and partially compensate each other. Unfortunately, both effects contain
large uncertainties. Indeed, the calculations \cite{milan2,milan3} were carried out using the Skyrme force
parameters, that are fixed  only in the vicinity of the Fermi surface, whereas the BCS gap equation involves the
function $m^*(k)$ for $k\gg k_{\rm F}$. The same is true for the induced interaction. In this case, the situation
is even worse since the spin channel in the high-frequency response function is important, but the corresponding
combinations of Skyrme force parameters are not known sufficiently well even at the Fermi surface. More detailed
analysis of these problems can be found in Ref. \cite{Bald1}. On the other hand, a similar compensation  between
self-energy and vertex corrections is  found in the calculation of pairing in nuclear matter\cite{lsz,cao},
supported with a Monte Carlo calculation of the gap\cite{MC}.

A ``blow-up'' was produced in 2008 by Duguet and Losinsky \cite{Dug1} who solved the {\it ab initio} BCS gap
equation for a lot of nuclei on the same footing. It should be noticed that the main difficulty of the direct
method of solving the nuclear pairing problem comes from the rather slow convergence of the sums over intermediate
states $\lambda$ in the gap equation, because of the short-range of the free $NN$-force. Evidently, this is the
reason why the authors of Refs. \cite{milan2,milan3} concentrated only on a single nucleus, i.e. $^{120}$Sn. To
avoid the slow convergence, the authors of Refs. \cite{Dug1,Dug2} used the ``low-k'' force $V_{\rm low-k}$
\cite{Kuo} which is in fact very soft. It is defined in such a way that it describes correctly the $NN$-scattering
phase shifts at momenta $k{<}\Lambda$, where $\Lambda$  is a parameter corresponding to
 the limiting energy  $E_{\rm lim} \simeq 300\;$MeV.  Moreover, the force $V_{\rm low-k}$ rapidly
vanishes  for $k{>}\Lambda$, so that in the gap equation one can
restrict the energy range
 to $E_{\max} {\simeq} 300\;$MeV . In addition a separable version of this force
was constructed that made it possible to calculate neutron and proton pairing gaps for a lot of nuclei. Usually
the low-k force is found starting from some realistic $NN$-potential ${\cal V}$ with the help of the
Renormalization Group method, and the result does not practically depend on the particular choice  of ${\cal V}$
\cite{Kuo}. In addition, in Ref. \cite{Dug1} $V_{\rm low-k}$ was found starting from the Argonne potential
v$_{18}$, that is different only a little bit from Argonne v$_{14}$, used in Ref. \cite{milan3}. Finally, in Ref.
\cite{Dug1} the same SLy4 self-consistent basis was used as in Ref. \cite{milan3}. Thus, the inputs of the two
calculations look very similar, but the results turned out to be strongly different. In fact, in Ref. \cite{Dug1}
the value $\Delta_{\rm BCS}\simeq 1.6\;$MeV was obtained for the same nucleus $^{120}$Sn which is already bigger
than the experimental one by $\simeq 0.3\;$MeV. In Refs. \cite{Bald1,Pankr1} we analyzed the reasons of these
contradictions. This point was discussed also in Ref. \cite{Dug2}. It turned out that, in fact, these two
calculations differ in the way they take into account the effective mass. Namely, in Refs. \cite{milan2,milan3}
the effective mass $m^*=m^*_{\rm SLy4}$ was used for all momenta up to $k_{\rm max}=\sqrt{2m^* E_{\rm max}}\simeq
6\;$fm$^{-1}$. On the other hand, in Refs. \cite{Dug1,Dug2} this mass  is used only for $k<\Lambda \simeq 3\;$
fm$^{-1}$, and the prescription $m^*=m$ is, in fact, imposed for bigger momenta. It implies that the gap $\Delta$
depends not only on the value of the effective mass at the Fermi surface, as it follows from the above exponential
formula for the gap, but also on the behavior of the function $m^*(k)$ in a wide momentum range. However, this
quantity is not known sufficiently well \cite{Bald1}, which makes rather uncertain the predictions of such
calculations.

To avoid such a drawback, we suggest a semi-microscopic model for
nuclear pairing containing a single phenomenological parameter. It
starts from the {\it ab initio} BCS gap equation with the Argonne
force v$_{18}$ treated with the two-step method. The complete
Hilbert space $S$ of the problem is split into the model subspace
$S_0$ of low-energy states and the complementary one $S'$. The gap
equation is solved in the model space with the effective interaction
$\Vef$ which is found in the complementary subspace. This {\it
ab-initio} term of $\Vef$ is supplemented by a small one-parameter
addendum that should hopefully embody all corrections to the
simplest BCS scheme with $m^*=m$. Preliminary results of this model
are reported in Ref. \cite{Pankr2}.

\section{Outline of the formalism}

We start from the general form of the many-body theory equation
for the pairing gap $\Delta$ \cite{AB},
%1
\beq \Delta_{\tau} =  {\cal U}^{\tau} G_{\tau} G^s_{\tau}
\Delta_{\tau}, \label{del} \eeq where $\tau=(n,p)$ is the isotopic
index, ${\cal U}^{\tau}$ is the $NN$-interaction block irreducible
in the two-particle $\tau$-channel, and
 $G_{\tau}$  ($G^s_{\tau}$) is the one-particle Green function without (with)
 pairing. A symbolic multiplication, as usual, denotes the integration over
energy and intermediate coordinates and summation over spin
variables as well. We have used above the term ``BCS theory''
meaning that, first, the block ${\cal U}$ of irreducible interaction
diagrams is replaced with the free $NN$-potential ${\cal V}$  in Eq.
(\ref{del}), and, second, the simple quasi-particle Green functions
for $G$ and $G^s$ are used, i.e. those without phonon corrections
and so on. In this case, Eq. (\ref{del}) turns greatly simplified
and can be reduced to the form usual in the Bogolyubov method, \beq
\Delta_{\tau} = - {\cal V}^{\tau} \varkappa_{\tau}\,, \label{delkap}
\eeq where \beq\varkappa_{\tau}=\int \frac {d\eps}{2\pi i}G_{\tau}
G_{\tau}^s\Delta_{\tau}
 \label{defkap}\eeq is the anomalous density matrix
which can be expressed explicitly in terms of the Bogolyubov
functions $u$ and $v$,
%3
\beq \varkappa_{\tau}({\bf r}_1,{\bf r}_2) = \sum_i u_i^{\tau}({\bf
r}_1) v_i^{\tau}({\bf r}_2). \label{kapuv} \eeq Summation in Eq.
(\ref{kapuv}) scans the complete set of Bogolyubov functions with
eigen-energies $E_i>0$.

As mentioned in the Introduction, we use a two-step renormalization
method of solving the gap equation in nuclei to overcome the slow
convergence problem. We split the complete Hilbert space of the
pairing problem $S$ to the model subspace $S_0$, including the
single-particle states with energies less than a fixed value of
$E_0$, and the subsidiary one, $S'$. The gap equation is solved in
the model space: \beq \Delta_{\tau} = \Vef^{\tau} G_{\tau}
G^s_{\tau} \Delta_{\tau}|_{S_0}, \label{del0} \eeq with the
effective pairing interaction $\Vef^{\tau}$ instead of the block
${\cal U}^{\tau}$ in the original gap equation (\ref{del}). It obeys
the Bethe--Goldstone type equation in the subsidiary space, \beq
\Vef^{\tau} = {\cal U}^{\tau} + {\cal U}^{\tau} G_{\tau} G_{\tau}
\Vef^{\tau}|_{S'}. \label{Vef} \eeq In this equation, the pairing
effects can be neglected provided the model space is sufficiently
large. That is why we replaced the Green function $G^s_{\tau}$ for
the superfluid system with its counterpart $G_{\tau}$ for the normal
system. In the BCS approximation, the block  ${\cal U}^{\tau}$ in
Eq. (\ref{Vef}) should be replaced by ${\cal V}^{\tau}$. To solve
this BCS version of Eq. (\ref{Vef}) in non-homogeneous systems, we
have found a new form of the local approximation, the Local
Potential Approximation (LPA). Originally, it was developed for
semi-infinite nuclear matter \cite{Bald0}, then for the slab of
nuclear matter (see review articles \cite{Rep,ST}) and, finally, for
finite nuclei \cite{Pankr1,Bald1}. It turned out that, with a very
high accuracy, at each value of the average c.m. coordinate ${\bf
R}=({\bf r}_1 + {\bf r}_2 +{\bf r}_3 +{\bf r}_4)/4$, one can use in
Eq. (\ref{Vef}) the formulae  of the infinite system embedded into
the constant potential well $U=U({\bf R})$. This explains the term
LPA, and also significantly simplifies the equation for $\Vef$, in
comparison with the initial equation for $\Delta$. As a result, the
subspace $S'$ can be chosen as large as necessary to achieve the
convergence. From the comparison of the direct solution of Eq.
(\ref{Vef}) in the slab with LPA, it was shown that LPA has high
accuracy, even in the surface region, for sufficiently large model
space, $E_0$ (${\simeq} 20{\div} 30\;$MeV). For finite nuclei,
including $^{120}$Sn, the validity of LPA was also checked
\cite{Pankr1,Bald1}. In this case, the boundary energy should be
made larger up to $E_0{=}40\;$MeV. In this paper, we use LPA with
this choice of $E_0$ for systematic calculations of the gap in
spherical nuclei. For ${\cal V}$, we use,  just as in Ref.
\cite{Bald1}, the Argonne potential v$_{18}$. To make the
calculations more treatable, we use the separable representation
\cite{Bal-sep} of the v$_{18}$ potential. Even in this simplified
version the calculations of the set of matrix elements of $\Vef$ for
a single nucleus require about 30 hours cpu with 50 processors of
the multi-processor system of the Kurchatov Institute.

Let us notice that the use of the low-k force $V_{\rm low-k}$ could be also interpreted in terms of the two-step
renormalization scheme of solving the gap equation (\ref{del}), with  $E_0 {\simeq} 300\;$MeV and with free
nucleon Green functions  $G$ in Eq. (\ref{Vef}) (i.e. $U(R)=0$). Then, (with ${\cal U}{\to} {\cal V}$) one obtains
$\Vef{\to} V_{\rm low-k}$ (see Ref. \cite{Kuo-Br} where the usual renormalization scheme is used to find $V_{\rm
low-k}$ instead of the Renormalization Group equation). Now, the comparison of the direct solution of the gap
equation (\ref{del}) (or (\ref{delkap}))  in Ref. \cite{milan3} with the Argonne $NN$-potential ${\cal V}$ and of
``renormalized'' equation (\ref{del0}) with $\Vef = V_{\rm low-k}$ shows that the difference appears because, in
the subsidiary subspace $S'$, the effective mass $m^*{\neq} m$ is used in the first case, and $m^*=m$ in the
second one. Thus, the resulting gap depends not only on the value of the effective mass at the Fermi surface, but
also on the behavior of the function $m^*(k)$ in a wide momentum range. This dependence was demonstrated
explicitly in Refs. \cite{Pankr1,Bald1}. The use of the SHF effective mass corresponding to the SLy4 force, or to
any other version of the Skyrme force, could hardly be accepted. Indeed, these effective forces were introduced
and fitted to describe systematically nuclear masses and radii. As a rule, the description of the single-particle
spectrum nearby the Fermi surface with Skyrme forces is rather poor, and it is expected not to be less poor at
those high momenta that are involved in the gap equation (\ref{del}). This point makes it tricky the problem of
determine the pairing gap completely from first principles, because the many-body theory(\ref{del}) contains, in
addition to the ``$k$-mass'' of the SHF method, the ``$E$-mass'' (inverse $Z$-factor) \cite{bg1,bg2,lsz}, that is
not sufficiently well known even in nuclear matter \cite{Bald1,lsz,cao}. The corrections to the BCS version of Eq.
(\ref{del}) include also the difference of the block ${\cal U}$ from the potential ${\cal V}$, mainly due to the
so-called induced interaction. The attempt in Ref. \cite{milan3} to determine the latter from the SLy4 force
together with the nuclear mean field looks questionable. Indeed, the SLy4 parameters were fitted to the nuclear
mass table data mainly related to  the scalar Landau--Migdal (LM) amplitudes $f,f'$. As to the spin amplitudes
$g,g'$, they remain practically undetermined in the SHF method. But the contribution of the spin channel to the
induced interaction is not smaller than that of the scalar one \cite{milan3}. Parameters $g,g'$ are well known
from the calculations of nuclear magnetic moments within the Finite Fermi Systems (FFS) theory \cite{BST} but, as
for the Skyrme parameters, only at the Fermi surface. However, the states distant from the Fermi surface are
important to calculate the induced interaction. The induced interaction for such states has only been determined
in nuclear matter within the microscopic Brueckner theory \cite{cao}.
 At last, let us imagine to get from
phenomenology the functions $m^*(k), Z(k)$ and all the LM amplitudes
far from the Fermi surface. Even in this case, the use of so many
phenomenological ingredients devalues significantly the {\it ab
initio} starting point, i.e. the free $NN$ potential ${\cal V}$ in
the pairing gap calculation.

Instead, we suggest to introduce in the effective pairing interaction a small phenomenological addendum which
embodies, of course approximately, all the corrections to the BCS scheme discussed above. The simplest ansatz for
it is as follows: \beq {\cal V}^{\tau}_{\rm eff}({\bf r}_1,{\bf r}_2,{\bf r}_3,{\bf r}_4) = V^{\rm BCS}_{\tau,{\rm
eff}}({\bf r}_1,{\bf r}_2,{\bf r}_3,{\bf r}_4) + \gamma^{\tau} C_0 \frac {\rho(r_1)}{\bar{\rho}(0)}\delta ({\bf
r}_1 - {\bf r}_2)\delta ({\bf r}_1 - {\bf r}_3)\delta ({\bf r}_2 - {\bf r}_4). \label{Vef1} \eeq Here   $\rho(r)$
is the density of nucleons of the kind under consideration, and $\gamma^{\tau}$ are dimensionless phenomenological
parameters. To avoid any influence of the shell fluctuations in the value of ${\rho}(0)$, the average central
density ${\bar{\rho}(0)}$ is used in the denominator of the additional term. It is averaged over the interval of
$r{<}2\;$fm. The first, {\it ab initio}, term in the r.h.s. of Eq. (\ref{Vef1}) is the solution of the BCS version
of Eq. (\ref{Vef}) (with ${\cal U} \to {\cal V}$) in the framework of the LPA method described above, with
$m^*{=}m$ in the subspace $S'$.

We will see below that a rather small value of the phenomenological
parameter $\gamma_n=\gamma_p\simeq 0.06$  is sufficient to produce
 the necessary effect of suppressing theoretical gaps predicted by
the {\it ab initio} calculation.  The smallness of the
phenomenological addendum to the effective
 interaction itself  is demonstrated in Fig. 1 where the
 localized ``Fermi average''
 effective interaction is drawn for $\gamma=0$ and $\gamma=0.06$ values for
 two heavy nuclei. In the mixed coordinate-momentum representation, this
 quantity is defined as
follows: ${\cal V}_{\rm eff}({\bf k}_1,{\bf k}_2,{\bf r}_1,{\bf
r}_2)\to {\cal V}^{\rm F}_{\rm eff}(R=r_1) \delta({\bf r}_1-{\bf
r}_2) \delta({\bf r}_1-{\bf r}_3) \delta({\bf r}_2-{\bf r}_4)$,
where \beq {\cal V}^{\rm F}_{\rm eff}(R)= \int d^3t {\cal V}_{\rm
eff}(k_1=k_2=k_{\rm F}(R),{\bf R}-{\bf t}/2,{\bf R}+{\bf t}/2),\eeq
with $k_{\rm F}(R)=\sqrt{2m(\mu-U(R))}$, provided $\mu-U(R)\ge 0$,
and $k_{\rm F}(R)=0$ otherwise. Here $\mu$ and $U(R)$ are the
chemical potential and the potential
  well, respectively, of the kind
of nucleons  under consideration. A similar quantity was considered
 before in the nuclear slab to visualize the effective
interaction properties \cite{Rep,EPI}. At a glance, the difference
between the interaction strengths for $\gamma{=}0$ and
$\gamma{=}0.07$ is negligible, but it produces noticeable effects in
the gap owing to the exponential dependence on the force  of the gap
discussed in the Introduction.

 \begin{figure}
\centerline {\includegraphics [width=80mm]{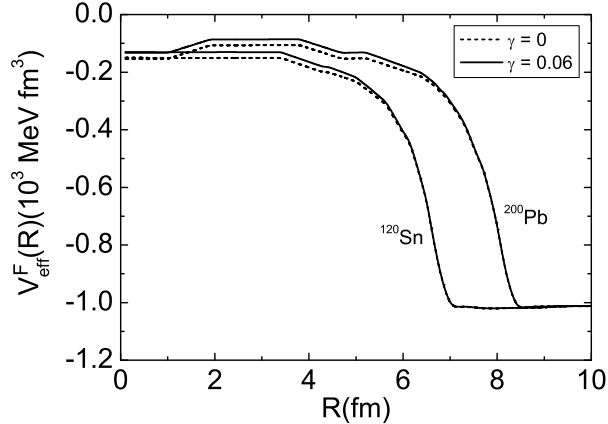}} \vspace{2mm}
\caption{The Fermi average effective pairing interaction ${\cal
V}^{\rm F}_{\rm eff}(R)$ for $^{120}$Sn and $^{200}$Pb nuclei, at
$\gamma=0$ (dashed curves) and $\gamma=0.076$ (solid curves)}
\end{figure}

In the case of proton pairing ($\tau=p$), we retain in Eq.
(\ref{Vef}) only the ``strong'' part of the pp potential ${\cal
V}^p$ and then add the ``bare''  Coulomb potential ${\cal V}_{\rm
C}$ to the BCS term of Eq. (\ref{Vef1}) , \beq {\cal V}_{p,\rm
eff}^{\rm BCS}={\cal V}^{\rm BCS}_{s,\rm eff}+{\cal V}_{\rm
C}.\label{Vefp}\eeq  Such approximation is valid because the mixed
strong-Coulomb term of Eq. (\ref{Vef}) is short-range, just as the
strong term itself, but it is proportional to the small parameter of
the fine structure constant $\alpha=1/137$. For matrix elements
$\langle\lambda_1\lambda_2| {\cal V}_{\rm
C}|\lambda_3\lambda_4\rangle$  of the bare Coulomb force this small
parameter is partially compensated in the diagonal case,
$\lambda_1=\lambda_2=\lambda_3=\lambda_4=\lambda$, owing to its
long-range character. Such matrix elements can be estimated as
$\simeq e^2/R$,  where $R$ is the nuclear radius. For example, for
$N=82$ isotonic chain it is of the order of 15-20\% of the main
diagonal matrix elements. It results in a$\simeq 30$\% suppression
of the proton gap value. So strong contribution of the Coulomb
interaction to the proton pairing was reported recently in Ref.
\cite{Dug2}. The above arguments to neglect the renormalization of
Coulomb interaction are valid also inside the model space. In
particular, the photon vertex ${\cal T}(q)$ does not change in the
long-range (small $q$) limit, owing to the Ward identity. In other
words, the large (long-range) Coulomb matrix elements are not
changed by the strong interaction. Small (short-range) matrix
elements do change, but they can be neglected. Thus, we can put
$\gamma_p=\gamma_n=\gamma$ in Eq. (\ref{Vef1}) after adding the bare
Coulomb interaction to the BCS term  for protons.

Then, the gap equation (\ref{del0}) in the model space is solved in the $\lambda$-representation,  with the
self-consistent basis determined within the Generalized Energy Density Functional (GEDF) method
\cite{Fay1,Fay2,Fay3,Fay4,Fay}, where $m^*{=}m$ is assumed, with the functional DF3 \cite{Fay4,Fay}. The latter is
of principal importance for our approach: first, because it makes the results less model-dependent, all effects of
$m^*\neq m$ in both model and subsidiary subspaces being attributed to the in-medium corrections beyond the pure
BCS approximation, and second, because single-particle spectra of the GEDF method  are, as a rule, in a better
agreement with experiment  than those of the popular versions of the SHF method \cite{HFB} (see for comparison
Ref. \cite{Tol-Sap}). The quality of the single-particle spectrum nearby the Fermi surface is very important for
obtaining the correct value of the gap calculated from Eq. (\ref{delkap}).

\section{Results}
 We solved the equations of the previous Section using the
self-consistent $\lambda$-basis of the GEDF method with the DF3 functional of Refs. \cite{Fay4,Fay}. The
discretization method for the continuum states was used in the spherical box of radius $R=16\;$fm with the grid
step $h=0.05\;$fm. The model space $S_0$ was extended up to the energy $E_0=40\;$MeV, the subsidiary one $S'$, up
to $E_{\rm max}=1000\;$MeV. The numerical stability of the results was checked by increasing the parameters up to
$E_0=60\;$MeV, $E_{\rm max}=1200\;$MeV and $R=24\;$fm, and we found for the gap value a numerical accuracy of 0.01
MeV.

In this paper, we limit ourselves to semi-magic nuclei. There are several reasons for such a choice. The first one
is of technical nature, namely we have only spherical code for the gap equation, and all or almost all semi-magic
nuclei are spherical. The second one is that in nuclei with both non-magic  subsystems there are often very
``soft'' low-lying $2^+$-states whose contributions to different quantities, in particular to the gap value, can
be rather strong and non-regular. In such cases, it is difficult to expect that the simple ansatz (\ref{Vef1})
will work with an universal parameter $\gamma$. For neutron pairing, we consider the lead, tin, and calcium
chains. The formulae above correspond to so-called ``developed pairing'' approximation \cite{AB}, that amounts to
impose the equality of the $\Delta^+$ and $\Delta^-$ operators and to neglect the particle-number non-conservation
effects. Therefore we limit ourselves to nuclei having, as a minimum, four particles (holes) above (below) the
magic core. For this reason, only the isotope $^{44}$Ca was considered in the calcium chain.

In accordance with the recipe of Ref. \cite{milan3}, we represent
the theoretical gap with the ``Fermi average'' combination \beq
\Delta_{\rm F}=\sum_{\lambda}{(2j{+}1)\Delta_{\lambda
\lambda}}/\sum_{\lambda}(2j{+}1), \label{DelF}\eeq where the
summation is carried out over the states $\lambda$ in the interval
of $|\eps_{\lambda}{-}\mu|{<}3\;$MeV, . The ``experimental'' gap is
determined by the usual 5-term mass difference, Eq. (\ref{dexp}) of
the Appendix. As it is argued in the Appendix, the relevance of the
mass difference to the gap has an accuracy of $\simeq(0.1
 \div 0.2)$ MeV. Therefore, it is reasonable to try to achieve
 the agreement of the gap within such limits.

Let us begin from the lead chain. Results are presented in Table I.
In the end of this
 table, the results  are shown for the $^{44}$Ca nucleus which is
  the lightest one among the scope ``from calcium to lead''
  considered in this article. We try to find a value of
  $\gamma$ which will be universal for all the region.

\begin{table}[t]
\caption{ Neutron gap $\Delta^n_{\rm F}$ (MeV) in Pb isotopes and
$^{44}$Ca nucleus.}

\begin{tabular}{c c c}
\hline \hline \hspace*{0.4ex} nucleus \hspace*{0.4ex} &
\hspace*{11.5ex} $\Delta^n_{\rm F}$
\hspace*{11.5ex} & \hspace*{0.7ex} $\Delta_{\rm exp}$\hspace*{0.7ex} \\
\end{tabular}

\begin{tabular}{c c c c c}
\hspace*{9.7ex} & \hspace*{1.5ex} $\gamma$=0 \hspace*{1.5ex} &
\hspace*{1.5ex} 0.06 \hspace*{1.5ex} & \hspace*{1.5ex} 0.08
\hspace*{1.5ex} &
\hspace*{3ex}  \hspace*{3ex} \\

\hline

$^{182}$Pb  & 1.79 & 1.33 & 1.20 & 1.30\\

$^{184}$Pb  & 1.79 & 1.33 & 1.20 & 1.34\\

$^{186}$Pb  & 1.78 & 1.32 & 1.19 & 1.30\\

$^{188}$Pb  & 1.76 & 1.31 & 1.17 & 1.25\\

$^{190}$Pb  & 1.73 & 1.29 & 1.16 & 1.24\\

$^{192}$Pb  & 1.68 & 1.22 & 1.09 & 1.21\\

$^{194}$Pb  & 1.62 & 1.16 & 1.03 & 1.13\\

$^{196}$Pb  & 1.53 & 1.09 & 0.96 & 1.01\\

$^{198}$Pb  & 1.43 & 1.00 & 0.87 & 0.94\\

$^{200}$Pb  & 1.31 & 0.90 & 0.80 & 0.87\\

$^{202}$Pb  & 1.16 & 0.79 & 0.69 & 0.78\\

$^{204}$Pb  & 0.95 & 0.64 & 0.56 & 0.71\\

\hline

$^{44}$Ca   & 1.83 & 1.50 & 1.41 & 1.54\\

\hline \hline
\end{tabular}\label{tab_Pb}
\end{table}

In Table I the strong effect of the small phenomenological addendum
in Eq. (\ref{Vef1}) is shown. The {\it ab initio} BCS result
($\gamma=0$) significantly overestimates the gap. Switching on this
term with $\gamma=0.06\div 0.08$ suppresses the gap
 by 30 - 40\%, in agreement with the data. The rms deviation of
 the theoretical values of the gap from the data for 13 nuclei, presented in Table
 I, is  $\sqrt{\overline{(\delta \Delta)^2}}{\simeq}0.045\;$MeV at
$\gamma=0.06$
 and 0.103 MeV at $\gamma=0.08$. It has a minimum $\sqrt{\overline{(\delta \Delta)^2}}
 {\simeq}0.037\;$MeV at $\gamma=0.064$, but, according to the
 above estimate, it is not reasonable to push the
 accuracy too much. In any case, we may consider the parameter $\gamma\simeq 0.06$
 as an optimal one for this set of nuclei.

Let us consider now the tin chain. The results are presented in
Table II.  Again we see a strong suppression with $\gamma=0.06\div
0.08$, but the agreement now is remarkably poorer than in the lead
case. Now, the rms deviation is $\sqrt{\overline{(\delta
\Delta)^2}}{\simeq}0.165\;$MeV at $\gamma=0.06$ and 0.169 at
$\gamma=0.08$, and the minimal value $\sqrt{\overline{(\delta
\Delta)^2}}{\simeq}0.158\;$MeV at $\gamma=0.07$ is also too large.

\begin{table}[t]
\caption{ Neutron gap $\Delta^n_{\rm F}$ (MeV) in Sn isotopes.}

\begin{tabular}{c c c }
\hline \hline \hspace*{0.4ex} nucleus \hspace*{0.8ex} &
\hspace*{11.0ex} $\Delta^n_{\rm F}$ &
 \hspace*{12.0ex} $\Delta_{\rm exp}$\hspace*{0.7ex} \\
\end{tabular}

\begin{tabular}{c c c c c}
\hspace*{9.7ex} & \hspace*{1.5ex} $\gamma$=0 \hspace*{1.5ex} &
\hspace*{1.5ex} 0.06 \hspace*{1.5ex} & \hspace*{1.5ex} 0.08
\hspace*{1.5ex} &
\hspace*{3ex}  \hspace*{3ex} \\

\hline
$^{106}$Sn  & 1.35 & 0.95 & 0.83 &  1.20\\

$^{108}$Sn  & 1.52 & 1.13 & 1.01 &  1.23\\

$^{110}$Sn  & 1.65 & 1.26 & 1.14 &  1.30\\

$^{112}$Sn  & 1.74 & 1.34 & 1.23 &  1.29\\

$^{114}$Sn  & 1.80 & 1.40 & 1.28 &  1.14\\

$^{116}$Sn  & 1.82 & 1.43 & 1.31 &  1.10\\

$^{118}$Sn  & 1.83 & 1.44 & 1.32 &  1.25\\

$^{120}$Sn  & 1.80 & 1.42 & 1.31 &  1.32\\

$^{122}$Sn  & 1.74 & 1.38 & 1.28 &  1.30\\

$^{124}$Sn  & 1.65 & 1.30 & 1.21 &  1.25\\

$^{126}$Sn  & 1.51 & 1.19 & 1.10 &  1.20\\

$^{128}$Sn  & 1.31 & 1.02 & 0.94 &  1.16\\

\hline \hline
\end{tabular}\label{tab_Sn}
\end{table}

Let us move now to protons. The effect of the Coulomb interaction to
the proton gap is shown  in Table III for the isotonic chain of
nuclei with the magic neutron number $N=82$. It is seen that,
indeed, it is rather strong $\simeq 0.5\;$MeV, in accordance with
\cite{Dug2}. Again at $\gamma {=}0.06$ the agreement is almost
perfect for the most part of nuclei, and only for the two heaviest
isotones the disagreement is of the order of 0.2 MeV. In this case,
a possible explanation lies in that we are close to the phase
transition to a deformed state (at $A\simeq 150$). Owing to the
contribution of these ``bad'' cases, the average difference between
the theoretical and experimental gaps for the $N=82$ chain is rather
high, $\sqrt{\overline{(\delta \Delta)^2}}=0.124\;$MeV. The average
error for all 34 nuclei considered is equal to
$\sqrt{\overline{(\delta \Delta)^2}}{\simeq}0.086\;$MeV. As it
follows from the analysis discussed in the Appendix, this value is
within the accuracy of the experimental values of the gap extracted
from the 5-term formula (\ref{dexp}).

\begin{table}[t]
 \caption{ Proton gap $\Delta^p_{\rm F}$
(MeV) for the isotone gap $N=82$.}
\bigskip

\begin{tabular}{c c c}
\hline \hline \hspace*{5.0ex} nucleus \hspace*{0.4ex} &
\hspace*{3.5ex} $\Delta^p_{\rm F}$
\hspace*{15.3ex} & \hspace*{0.7ex} $\Delta_{\rm exp}$\hspace*{0.7ex} \\

\end{tabular}

\begin{tabular}{c c c}
\hspace*{17.0ex} ${\cal V}_{\rm eff}^p{=}{\cal V}_{\rm eff}^0$ &
\hspace*{3.0ex}
${\cal V}_{\rm eff}^p{=}{\cal V}_{\rm eff}^0 + {\cal V}_{\rm C} $ \hspace*{6.ex} & \hspace*{7.ex} \\

\end{tabular}

\begin{tabular}{c c c c c }
\hspace*{9.7ex} & \hspace*{3ex}  \hspace*{3ex} & \hspace*{1.5ex}
$\gamma$=0 \hspace*{1.5ex} & \hspace*{1.5ex} 0.06 \hspace*{1.5ex}&
\hspace*{3ex}  \hspace*{3ex} \\

\hline

$^{136}$Xe  & 1.65 & 1.19 & 0.87  & 0.75\\

$^{138}$Ba  & 1.80 & 1.33 & 0.98  & 0.87\\

$^{140}$Ce  & 1.90 & 1.42 & 1.03  & 0.97\\

$^{142}$Nd  & 1.99 & 1.48 & 1.06  & 1.00\\

$^{144}$Sm  & 2.01 & 1.49 & 1.05  & 1.02\\

$^{146}$Gd  & 2.02 & 1.50 & 1.05  & 1.13\\

$^{148}$Dy  & 2.01 & 1.50 & 1.06  & 1.19\\

$^{150}$Er  & 1.98 & 1.48 & 1.07  & 1.22\\

$^{152}$Yb  & 1.92 & 1.44 & 1.05  & 1.29\\

\hline \hline

\end{tabular}\label{tab_P}
\end{table}

\section{The role of the single-particle spectrum}

We consider the poor agreement for the tin chain as a troubling point. Indeed, this chain is a traditional
benchmark for the pairing problem in nuclei, and the most strong deviation takes place for the $^{116}$Sn nucleus
which is in the very center of the chain where the scheme used should work especially well. In searching the
reasons for such a drawback, we paid attention to a new version of the DF3 functional, named DF3a \cite{Tol-Sap},
in which the spin-orbit and effective tensor terms of the original DF3 functional were changed to fit new data on
spin-orbit splitting in magic nuclei \cite{e_exp}. The matter is that the details of neutron paring in the chain
under consideration depend essentially on the position of the ``intruder'' state $1h_{11/2}$ which, in turn, is
determined mainly by these components of the functional. We repeated the calculations for this functional DF3a and
found that the agreement became much better. The optimal value $\gamma=0.052$ with $\sqrt{\overline{(\delta
\Delta)^2}}{\simeq}0.056\;$MeV is now a little less than for the lead chain, but for $\gamma=0.06$ chosen above
the agreement is also reasonably good, $\sqrt{\overline{(\delta \Delta)^2}}{\simeq}0.084\;$MeV. In Fig. 2 is
displayed the comparison of the theoretical predictions for the pairing gap in the tin chain (both versions of the
functional under consideration) with the experimental data.

\begin{figure}
\centerline {\includegraphics [width=80mm]{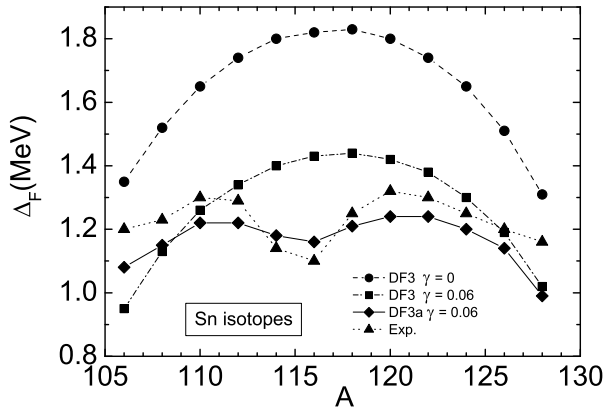}} \vspace{2mm}
\caption{Neutron gap in Sn isotopes }
\end{figure}

We see that, indeed,  the new calculation is now in nice agreement
with the data. To find the reason of so strong difference of the
results for these versions of the essentially same functional, we
examined the corresponding single-particle spectra along the chain
comparing them with existing experimental data. Dealing with an even
$^{A}$Sn isotope, there is a dilemma how to consider a state $|i
\rangle$ under consideration, either the hole state (i.e. the
excitation of the $^{A-1}$Sn nucleus) or the particle one (the
excitation of the $^{A+1}$Sn nucleus). We use a simple recipe: the
state $|i \rangle$ is considered as a hole state if the inequality
$v_i^2>0.5$ takes place and as a particle state otherwise. Note that
in the case of  $v_i^2 \simeq 0.5$ the difference between the
particle and hole energies is, as a rule, quite small. In general,
both the functionals reproduce the experimental low-lying  levels
sufficiently well. In particular, the spin of the ground state of
odd isotopes is always reproduced correctly for both  calculations.
But there is a noticeable difference for the
 $1h_{11/2}$ state in the left part of the chain, till  $^{122}$Sn, see Fig. 3.

\begin{figure}
\centerline {\includegraphics [width=120mm]{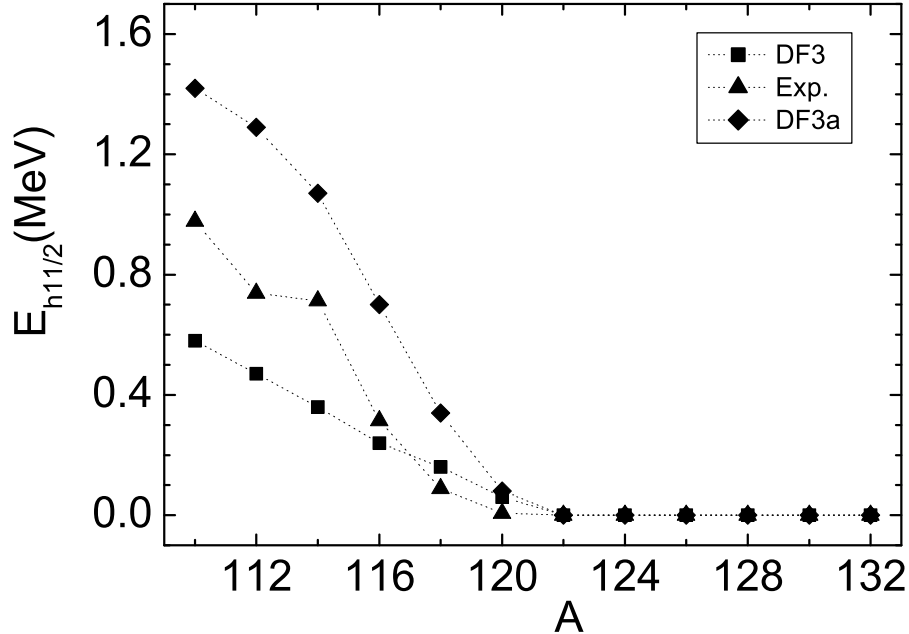}} \vspace{2mm}
\caption{The $h_{11/2}$ level position accounted from the ground
state in Sn isotopes}
\end{figure}

In this region, its position is  systematically lower than the
experimental one for the DF3 functional and  systematically higher,
for the DF3a functional.
 In the vicinity of the $^{116}$Sn nucleus, this
 difference becomes  dramatic and the too low position of the
 $1h_{11/2}$ state in the DF3 case leads to an incorrect enhancement of
 the gap value.

 In Fig. 4, the single-particle spectrum of the $^{114}$ Sn nucleus
calculated for the two versions of the functional is compared with the experimental one. Each theoretical level is
supplemented with the $(2j_i+1)u_iv_i$ factor that determines mainly the contribution of the $i$-level to the gap
equation. The same for $^{116}$Sn nucleus is displayed in Fig. 5. We see that for all other states $|i \rangle$
these factors are rather close for the two versions of the functional, but for the $1h_{11/2}$ state the
difference is rather strong. As to heavy tin isotopes for which this level becomes the ground state for both
functionals, the difference between their predictions for the gap value are quite close, see Fig. 2.

\begin{figure}
\centerline {\includegraphics [width=80mm]{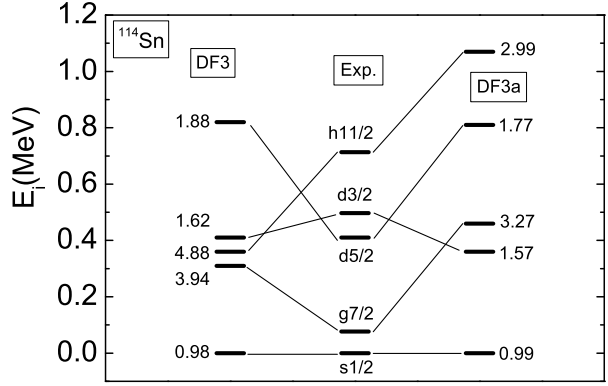}} \vspace{2mm}
\caption{$^{114}$Sn spectrum}
\end{figure}

\begin{figure}
\centerline {\includegraphics [width=80mm]{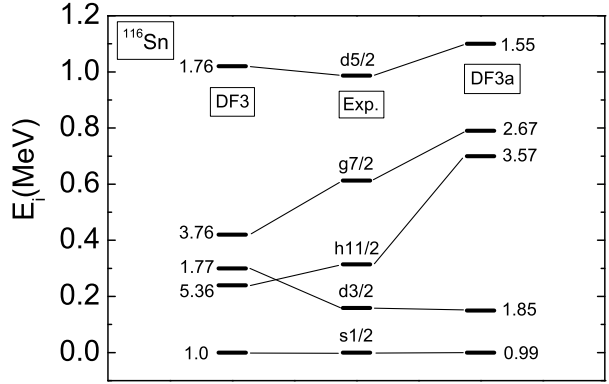}} \vspace{2mm}
\caption{$^{116}$Sn spectrum}
\end{figure}

 This analysis shows the great sensitivity of the gap value to the
single-particle spectrum nearby the Fermi level, especially to the
position of levels with high $j$-value. Therefore it is interesting
to examine which effect is to be expected in the other cases of
going from the initial DF3 functional to this new version. It is
displayed in Fig. 6 for the lead isotopes. We see that in this case
the overall agreement for the new version of the functional becomes
worse. Evidently, again the position of high $j$-levels is
different, in favor of the DF3 functional in this case.

\begin{figure}
\centerline {\includegraphics [width=100mm]{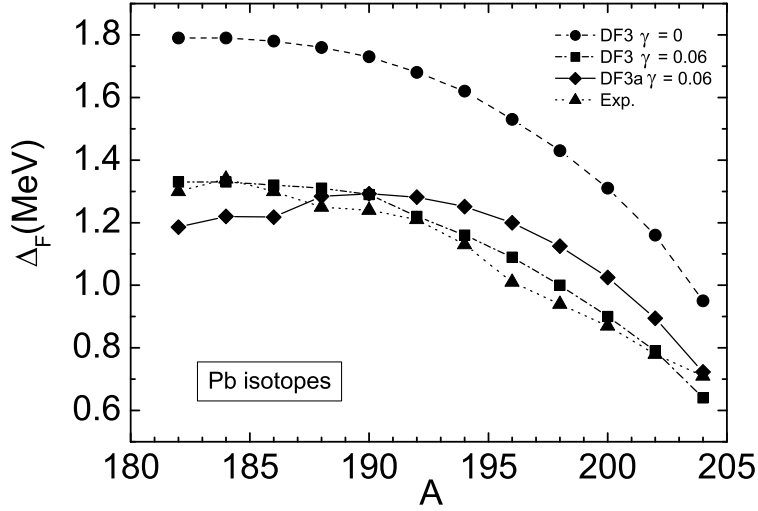}} \vspace{2mm}
\caption{Neutron gap in Pb isotopes. }
\end{figure}
A bad situation for the DF3a functional rises for the $N=82$ isotone
chain, see Fig. 7. The analysis shows that again the $1h_{11/2}$
level is guilty, now for protons. For the DF3a functional it is
again higher than the experimental position, and for $^{144}$Sm and
$^{146}$Gd nuclei much higher. As the result, it does not
practically contribute to the gap equation resulting in a too small
gap value.

\begin{figure}
\centerline {\includegraphics [width=100mm]{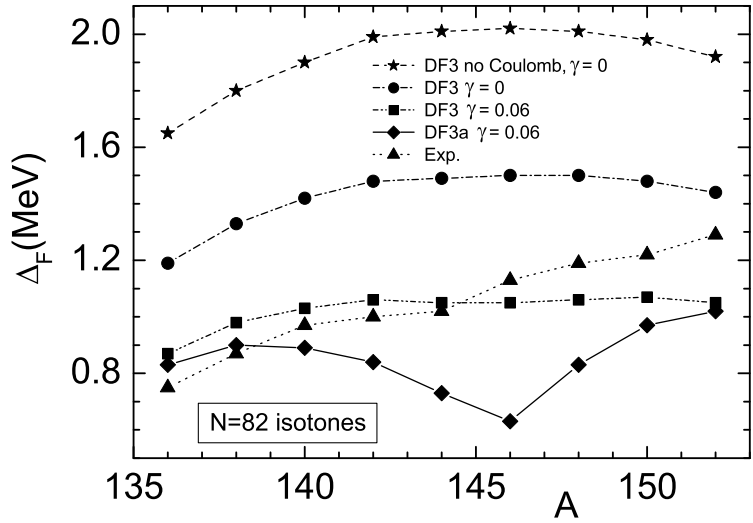}} \vspace{2mm}
\caption{Proton gap for $N=82$ isotones}
\end{figure}

\vskip 1cm
\section{Conclusion}
 We suggest a simple semi-microscopic model for the nuclear pairing
starting from the {\it ab initio} BCS gap equation. The
self-consistent GEDF Method basis, characterized by the bare nucleon
mass, is employed in the calculation. The gap equation is recast in
the model space $S_0$, replacing  the bare interaction with the
effective pairing interaction determined in the complementary
subspace $S'$.

The Argonne v$_{18}$ potential was adopted along with the LPA
method. A small phenomenological term is added to this effective
interaction that contains one parameter which should embody
approximately the effective mass and various corrections to the pure
BCS theory. Calculations are carried out with the DF3 functional
\cite{Fay4,Fay} for semi-magic lead and tin isotopic chains and the
$N=82$ isotonic chain as well. The Coulomb interaction is explicitly
included in the proton gap equation. We find that the model
reproduces rather well the experimental values of the neutron and
proton gaps in semi-magic nuclei. The overall agreement
($\sqrt{\overline{(\delta \Delta)^2}}{\simeq}0.13\;$MeV) is better
than that obtained in Ref. \cite{Dug1}, where the authors did not
introduce free parameters explicitly but they made it implicitly by
using a specific $k$-dependence in the effective mass.

We examine also the role of the single-particle spectrum in the gap equation. For this aim we use the new
modification DF3a \cite{Tol-Sap} of the functional \cite{Fay4,Fay} that changes spin-orbit and effective tensor
terms. The use of this functional gives an  agreement better for the tin chain and worse for the lead chain and
even more for the $N=82$ chain. The accuracy of the predictions depends strongly on the quality of reproducing the
positions of high $j$-levels in the self-consistent basis used. We are thinking, e.g., to the $1h_{11/2}$ neutron
level in the tin isotopes and $1h_{11/2}$ proton level for $N=82$ isotones.

 The ansatz of Eq. (\ref{Vef1}) exhibits an
obvious drawback. The phenomenological GEDF pairing interaction of Ref. \cite{Fay} contains the surface term
(${\propto}(d\rho /dr)^2$) that plays an essential role for the description of the odd-even effect (staggering) in
nuclear radii. It originates mainly from the exchange by surface phonons which was explicitly taken into account
in \cite{milan2,milan3}. The addition, to such a term in Eq. (\ref{Vef1}) is associated the introduction of a new
parameter, and at the first stage we prefer   to avoid that. A more consistent scheme should, evidently, include
the explicit consideration of the low-lying phonons, as e.g. in \cite{milan2}, but  taking into account the
so-called tadpole diagrams \cite{Kam_S}. In this case, the phenomenological constant $\gamma$ should, of course,
change.

\par
\noindent{\bf Acknowledgments}\par
 We thank G. L. Colo, T. Duguet, V. A. Khodel and S. V. Tolokonnikov for
valuable discussions. This research was partially supported by the joint Grants of RFBR and DFG, Germany, No.
09-02-91352-NNIO\_à, 436 RUS 113/994/0-1(R), by the Grants NSh-7235.2010.2  and 2.1.1/4540 of the Russian Ministry
for Science and Education, and by the RFBR grants 09-02-01284-a, 09-02-12168-ofi\_m. Two of us (S. P. and E. S.)
thank the INFN, Sezione di Catania, for hospitality.
%\end{acknowledgments}

\appendix

\section{Accuracy of extracting experimental gap values from the mass differences}

 In this Appendix, we discuss the accuracy of determination of
 the ``experimental'' gap,
 $\Delta_{\rm exp}$, from the mass data.
 Usually, this quantity is found in terms of
 mass values $M$ of neighboring nuclei via 3-term formulae,
 \beq 2\Delta^+_{\rm exp}(A)= \delta_2M^+ \equiv
 2M(A+1)-M(A+2)- M(A)\label{dexp1}\eeq or
\beq 2\Delta^-_{\rm exp}(A)= \delta_2M^- \equiv 2M(A-1)-
M(A-2)-M(A).\label{dexp2}\eeq The 5-term expression is usually
considered  more accurate, being a half-sum of them, \beq
\Delta_{\rm exp}(A)= \overline{\delta_2 M}/2
  \equiv (\delta_2M^+ + \delta_2M^-)/4.
\label{dexp}\eeq These simple recipes were used, in particular, in
\cite{milan2,milan3,Dug1,Dug2}. However, they originate from the
simplest model of $\Delta={\rm const}$, and  the accuracy of such
prescription  is not {\it a priori} obvious. To clarify this point
we made a calculation which could be considered as a ``theoretical
experiment''. We used the GEDF method \cite{Fay} with the functional
DF3 which reproduces the mass differences of Eqs. (\ref{dexp1}),
(\ref{dexp2}) type sufficiently well. We calculated first directly
the right side of Eq. (\ref{dexp}) and second, the theoretical gap
value with Eq. (\ref{DelF}) within the same GEDF method. The
comparison of these two quantities  is given in Fig.~7 for the lead
isotopes and in Fig.~8 for the tin isotopes. We see that for the
main part of nuclei under consideration the difference between
values in two neighboring columns is within 0.1 MeV. However, there
is several cases where it is of the order (or even exceeds) 0.2 MeV.
Leaving aside detailed analysis of these ``bad'' cases we are forced
to put a limit of $\simeq 0.1 - 0.2\;$MeV for the accuracy of the
experimental gap determined from Eq. (\ref{dexp}).

\begin{figure}
\centerline {\includegraphics [width=100mm]{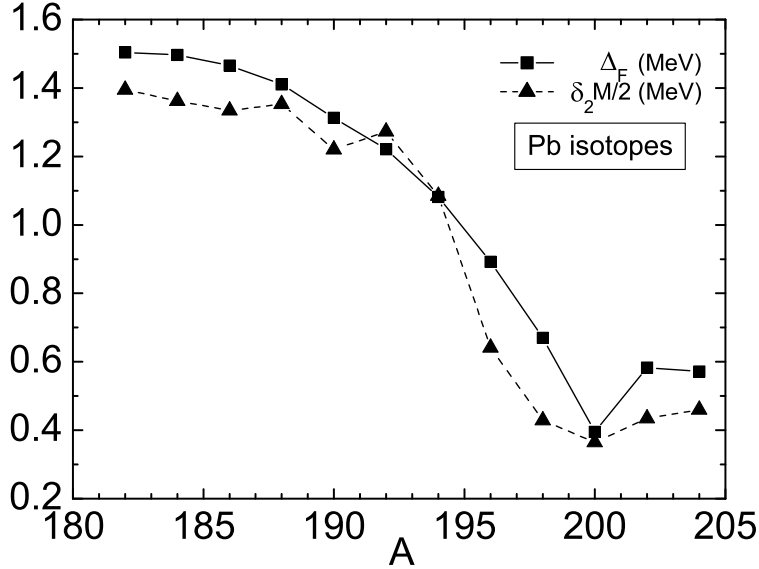}} \vspace{2mm}
\caption{The theoretical predictions for the mass difference
$\overline{\delta_2 M}/2 $ (triangles) versus the average gap value
$\Delta_{\rm F}$ (squares) for Pb isotopes. Both the quantities are
calculated within the GEDF method, see the text.}
\end{figure}

\begin{figure}
\centerline {\includegraphics [width=100mm]{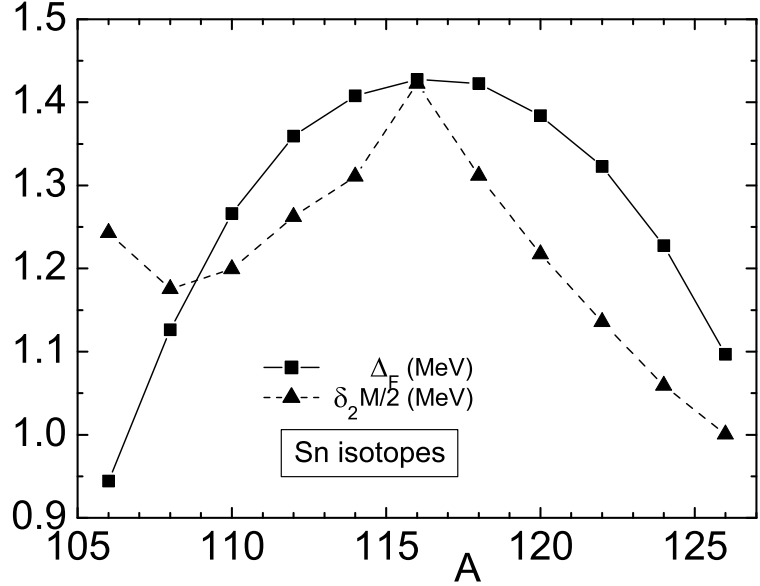}} \vspace{2mm}
\caption{The same as in Fig. 1, but for Sn isotopes.}
\end{figure}

{}
\end{document}